\documentclass[a4paper,twocolumn,11pt]{quantumarticle}
\pdfoutput=1
\usepackage[utf8]{inputenc}
\usepackage[english]{babel}
\usepackage[T1]{fontenc}
\usepackage{amsmath}
\usepackage{hyperref}
\usepackage{amssymb}

\usepackage{tikz}
\usepackage{lipsum}

\usepackage{ltxgrid}

\begin{document}

\title{Pretty good quantum state transfer on isotropic and anisotropic
Heisenberg spin chains with tailored site dependent exchange couplings}

\author[1]{Pablo Serra }

\author[2]{Alejandro Ferr\'on }

\author[1]{Omar Osenda }

\affiliation[1]{
 Instituto de Física Enrique Gaviola, Universidad Nacional de Córdoba
, CONICET, Facultad de Matemática, Astronomía, Física y Computación,
Av. Medina Allende s/n, Ciudad Universitaria, CP:X5000HUA Córdoba, Argentina.}

\affiliation[2]{
 Instituto de Modelado e Innovación Tecnológica (CONICET-UNNE) and
Facultad de Ciencias Exactas, Naturales y Agrimensura, Universidad Nacional
del Nordeste, Avenida Libertad 5400, W3404AAS Corrientes, Argentina.}

\maketitle

%%%%%%%%%%%%%%%%%%%%%%%%%%%%%%%%%%%%%%%%%%%

\begin{abstract}

Using a global optimization algorithm we obtain spin chains with site-dependent
exchange coefficients which allow almost perfect quantum state transfer between
the extremes of the chains without any further time-dependent external
control. We consider chains with isotropic and anisotropic Heisenberg
Hamiltonian with up to 100 spins. The method allow us to choose the arrival
of the transferred state by changing the range available to the exchange coupling strengths. We consider short transferred times, in particular shorter than
those achievable with known time-dependent control schemes. The chains
obtained with the optimization method show  some remarkable and interesting
traits as, for instance, the scaling   of the magnitude of the exchange
couplings needed to achieve near perfect state transfer with the length of
the chain and the arrival time. This scaling makes it possible to decide if
the chain with site-dependent coefficients can be implemented in an actual
system according to the range of interactions allowed in it. We compare the
robustness of the transmission protocol against static disorder on the exchange coefficients using different figures of merit, which allow us to conclude that
the isotropic Heisenberg chain is the best option when compared with
anisotropic chains. This comparison is relevant since the method allow us to
start with chains that all achieve near perfect quantum state transfer, regime
that was not available previously in these chains without time-dependent
external control.
\end{abstract}
%\pacs{73.22.Pr, 73.43.Cd}

\section{Introduction}\label{sec:introduction}

The transfer of quantum state along communication channels has been under study 
for almost two decades now as a task of primordial importance to Quantum 
Information Processing \cite{Nikolopoulos2015}. The cornerstone of the subject 
on quantum spin chains 
was laid by Bose \cite{Bose2003,Bose-review}, who introduced the figure of merit 
to analyze 
the quality of the resulting transfer processes.

The most simple transfer protocol on a spin chain can be described rather 
easily: a given state is prepared at one extreme of the chain and after some 
time another state, the transferred one,  is retrieved at the other extreme. 
The transfer is better for larger values of the fidelity, and perfect if the 
state retrieved is the same that the one that was prepared, which corresponds to 
the maximum value allowed to the fidelity. 

The dynamical behaviour between the  
preparation and retrieval times can be controlled but, in many cases, it is 
preferred that the time evolution of the quantum state is governed only by the 
time-independent spin chain Hamiltonian, which depends only on the exchange 
couplings between the spins on the chain. The unforced dynamics is preferred 
under the rationale that the precise tuning of the control fields applied to 
the quantum spin chain add other sources of noise that spoil the state 
transfer process \cite{Christandl2004,Christandl2005,Zwick1,Zwick2,Zwick3}. 
However, the strength of the exchange couplings can be modulated to improve the 
quality of the transfer process, resulting in a number of strategies and 
dynamical regimes \cite{ 
Christandl2004,Christandl2005,Zwick1,Zwick2,Zwick3,Yung2006,Burgarth2005,
Burgarth2005b,Banchi2010,Banchi2011}

The 
addition of noise sources is problematic enough but it is not the only drawback 
in actual implementations, since the strength of the coupling between the spin 
chain and the control fields and the frequency 
spectrum available are also causes for concern \cite{Wang2010}. 

Despite what has been said  
above, much progress has been made in the field of controlled state transfer, 
from the study of the  most basic aspects of controllability 
\cite{Jurdjevic1972,Burgarth2009,Wang2016,Ramakrishna1995}, to obtaining 
particular states in different types of chains  
\cite{Watanabe2010,Burgarth2010,Heule2010,Stefanatos2019}. There is a host of 
physical systems where some quantum state transfer process can be implemented 
\cite{Kostak2007,quantum-dot-chain,Li2018,nuclear-spin-chain,Loft2011,
Banchi2011prl,Chapman2016,Kandel2019}. The strategies to control a given chain 
are also multiple, varying from using only one actuator 
\cite{Wang2016,Burgarth2010,Yang2010}, two actuators \cite{Zhang2016} or even 
controlling almost all (or all) the spins on the chain 
\cite{Gong2007,Murphy2010,Faroq2015}.

On the other hand, perfect state transfer without other interactions than the 
exchange couplings between the spins of the chain can only be achieved in 
chains with the so called XX Hamiltonian \cite{Christandl2005}. The Heisenberg 
Hamiltonian only 
admits what has been termed {\em pretty good} 
\cite{Kay2010,Kay2019,Banchi2017,vanBommel2010} or near perfect state transfer, 
{\em i.e.} choosing adequate values for the exchange couplings strength it is 
possible to obtain fidelity values differentially close to unity. This scenario 
can be rigorously proved for chains whose length is of the form $N=2^k$, where 
$k$ is some natural number. It is interesting that these results can be proved 
using linear algebra theorems that provide insights about the Hamiltonian 
spectrum. 

Regrettably, the theorems that assure that the pretty good state transfer 
regime exists do not provide  neither the order of magnitude of the 
couplings nor the transmission time at which to look for them. Taking this into 
account, the optimization of the fidelity as a function of the exchange 
couplings strength and the transfer time can offer a different approach to the 
problem. There has been  attempts in this direction, for instance in 
Ref.~\cite{Zhang2018}  machine learning is employed to obtain the 
exchange couplings of the XX Hamiltonian that is associated to perfect state 
transfer. We are interested in calculate a set of parameters of the Hamiltonian 
that places us in the pretty good transfer regime. That is, we do not require 
the best set, but a very good set of parameters. The problem will be written as
the finding of a set of parameters that produce a (very good local) maximum in a 
cost function.

These class of problems, where we look for good maximum   of an arbitrary cost 
function, are suitable to be treated by stochastic optimization methods. In 
particular, in this paper 
we use a method developed by one of us, which has been shown to be very 
efficient in optimizing functions in parameter space of dimensions even greater 
than those of interest  in this study: the {\it pivot method } (PM) for global 
optimization \cite{ssk97,sskb97}.  The {\it pivot method} is a 
stochastic method to search for the global minimum of a multiple-minimum 
function defined on a high-dimensional phase space.
To find the  global minimum of a  high-dimensional corrugated landscape is, in 
general, a NP-hard problem, and therefore there is not a stochastic method that 
can  assure that the global minimum is reached.
However, the pivot method has shown to be an efficient method to find a very 
good minimum, that actually is what we need in the search for good quantum 
state transfer in chain-spin systems described in this paper.

The paper is organized as follows, In Section~\ref{sec:model-hamiltonians} the 
model Hamiltonians to be considered are introduced together with the transfer 
protocol and the figures of merit that quantify the quality of the transfer 
process, in particular the relationship between the fidelity of transmission 
and the {\em transferred population}, which is the probability that a single 
excitation is transferred from one end of the chain to the other. In this Section
we also describe how the optimization algorithm must be implemented to find the site-dependent exchange couplings that maximize the transferred population for a chosen transferral time, at the same time we introduced all the notation related to the numerical procedure.

In
Section~\ref{sec:Heisenberg} the transfer properties of isotropic Heisenberg 
spin chains with modulated  exchange coupling coefficients are thoroughly 
analyzed. Particular emphasis is placed on obtaining chains that transmit 
states with high fidelity for different transfer (or arrival) times.  It is 
shown that the magnitude of the largest exchange coefficient of a modulated 
chain depends on the arrival time and an approximate formula that fits the 
dependency  is proposed. It is studied how the transmission of a chain whose 
exchange coefficients have been optimized  is affected by the presence of 
static disorder. It is shown that the transferred population shows
scaling properties as a function of time when comparing the transfer of chains, 
of fixed length, but whose coefficients have been optimized for different 
arrival times. We also present a brief discussion about the
quality of the transfer protocol if the initial state can not be prepared exactly
like the protocol prescribes. Once the relationship between the chain length and
the maximum value of the exchange coefficients needed to achieve near perfect
quantum state transfer at a given arrival time is established, we discuss in which
kind of physical system  it could be possible to implement chains like the ones
treated in the Section.

The analysis described in the paragraph above is repeated in
Section~\ref{sec:XXZ-Hamiltonian} for anisotropic Heisenberg spin chains.  
The conclusions and the discussion of our results are presented in 
Section~\ref{sec:conclusions} where we discuss the different dynamical behaviours
that appear in the disordered spin chains.

\section{Model Hamiltonians, transfer protocol and Methods}\label{sec:model-hamiltonians}

The XXZ Hamiltonian of a quantum spin chain with only nearest-neighbor 
interactions can be written as
\begin{equation}
 \label{eq:XXZ-Hamiltonian}
 H = - \sum_{i=1}^{N-1} J_i \left( \sigma_i^x \sigma_{i+1}^x + \sigma_i^y 
\sigma_{i+1}^y + \Delta  \sigma_i^z \sigma_{i+1}^z \right) ,
\end{equation}

\noindent where $N$ is the number of spins, or the length of the chain,  
the $\sigma$'s are the Pauli matrices, the $J_i$'s  are the exchange coupling 
strengths, $J_i>0, \forall i$,  and $\Delta$ is an anisotropy parameter. The 
Heisenberg Hamiltonian correspond to the isotropic case, $\Delta=1$ and 
$\Delta=0$ is known as the XX Hamiltonian. Note that 
the  Hamiltonian in Eq.~\eqref{eq:XXZ-Hamiltonian} is adimensional, besides we 
take $\hbar=1$.

The 
Hamiltonian in Eq.~\eqref{eq:XXZ-Hamiltonian} commutes with the total 
magnetization in the $z-$direction

\begin{equation}
 \left[ H, \sum_i \sigma_i^z \right] = 0,
\end{equation}

\noindent so the Hamiltonian can be diagonalized in subspaces with fixed number 
of excitations, {\em i.e.} in subspaces with a given number of spins up. It is 
customary to use the computational basis, where for a single spin $|0\rangle 
=|\downarrow\rangle$ and $|1\rangle =|\uparrow\rangle$, so
$|\mathbf{0}\rangle = 
|0000\ldots 0\rangle$ is the state with zero spins up of the whole chain, and 

\begin{equation}
 \sum_{i=1}^N \sigma_i^z |\mathbf{0}\rangle = -N |\mathbf{0}\rangle.
\end{equation}

The $N$ states with a single spin up are denoted as follows
\begin{equation}
|\mathbf{1} \rangle = |10\ldots 0\rangle, |\mathbf{2} \rangle = 
|010\ldots0\rangle,\ldots , |\mathbf{N} \rangle = |00\ldots 1\rangle.
\end{equation}
 
\noindent {\em i.e.} the state $|\mathbf{j} \rangle $ is the state of the chain 
with only the $j$-th spin up. The one-excitation basis plays a fundamental role 
in the simplest transfer protocol. In this protocol the initial state of the 
chain, $|\Psi(0)\rangle$, is prepared as 

\begin{equation}\label{eq:initial-chain-product}
|\Psi(0)\rangle =|\psi(0)\rangle \otimes |\mathbf{0}\rangle_{N-1}
\end{equation}

\noindent where $|\psi(0)\rangle = \alpha |0\rangle +\beta |1\rangle $, is an 
arbitrary 
one-spin pure state, with $\alpha$ and $\beta$ complex constants such that 
$|\alpha|^2 + |\beta|^2 =1$, and $|\mathbf{0}\rangle_{N-1}$ is the state 
without excitations of a chain with $N-1$ spins. The state in 
Eq.~\eqref{eq:initial-chain-product} can be rewritten as

\begin{equation}
 \label{eq:initial-state}
|\Psi(0)\rangle = \alpha |\mathbf{0}\rangle + \beta |\mathbf{1} \rangle . 
\end{equation}
Using the time evolution operator
\begin{equation}
\label{eq:time-evolution}
U(t) = \exp{(-i H t)},
\end{equation}
the state of the chain at time $t$ can be obtained as
\begin{equation}
 \label{eq:time-dependent-state}
 |\Psi(t)\rangle = U(t) |\Psi(0)\rangle = \alpha U(t) |\mathbf{0}\rangle + 
\beta U(t) |\mathbf{1}\rangle .
\end{equation}
It is clear that the state transfer is perfect when for some time $T_{per}$ the state 
of the chain is given by
\begin{equation}
\label{eq:perfect-transfer}
|\Psi(T_{per})\rangle = e^{i\theta}\left(\alpha |\mathbf{0}\rangle + \beta 
|\mathbf{N}\rangle\right),
\end{equation}
where $\theta$ is some real arbitrary phase. So, the transfer fidelity as a 
function of time is calculated as

\begin{equation}\label{eq:first-fidelity}
 f_{\alpha,\beta}(t) = |\alpha \langle \Psi(t)|\mathbf{0}\rangle + \beta 
\langle \Psi(t)
|\mathbf{N}\rangle|^2 .
\end{equation}

The fidelity in Eq.~\eqref{eq:first-fidelity} depends on the particular initial 
state, Eq.~\eqref{eq:initial-state}, chosen to be transferred. Usually, to 
quantify the quality of the state transfer protocol, it is introduced the 
average of  $f_{\alpha,\beta}(t) $ over all the possible values of $\alpha$ and 
$\beta$. It can be shown that this averaged fidelity, $F(t)$, can be written as

\begin{equation}\label{eq:fidelity}
 F(t) = \frac{\sqrt{|P(t)|}}{3} + \frac{P(t)}{6}+\frac12 ,
\end{equation}

\noindent where

\begin{equation}\label{eq:population}
P(t) = |\langle \mathbf{1}|U(t)|\mathbf{N}\rangle|^2
\end{equation}

\noindent is 
called the 
 transferred population between the first and last sites of the 
chain,  and we assume that $\cos(\gamma)=1$, where 
$\gamma=arg(f_{\alpha,\beta}(t))$,  as in Ref.~\cite{Bose2003}. The 
pretty good transfer occurs when

\begin{equation}
 P(t) = 1- \epsilon, 
\end{equation}
where $\epsilon\ll 1$.\\

{\em The optimization algorithm}\\

The XX Hamiltonian shows perfect state transfer for a number of Exchange 
Couplings Distributions (ECDs). For ECD we understand a particular set of values
$\lbrace J_i \rbrace_{i=1}^{N-1}$ that univocally determines a particular
Hamiltonian for a chain. For example, it is well known that the ECD given 
by $J_i =\sqrt{i(N-i)}$ determines the XX Hamiltonian that shows perfect state
transfer at time $T_{per}=\pi/2$.

For Heisenberg chains $T_{per}$ it is not known, so the optimization method can be designed to obtain
an ECD for a chosen transfer time.

 Our method allow us to choose
the transfer time $T$, however we 
will focus on the range $T \in \left[\frac{N}{2},5N\right]$. The range chosen 
allows us to compare our results  for the Heisenberg Hamiltonian with 
those that are well known for the XX Hamiltonian. There is not an obvious upper 
value for the transfer times to be explored other than the search of fast 
transfer times. It is worth to mention that transfer times on the order $10N$ 
are achievable controlling the dynamics by applying  external magnetic fields 
 to the spins at the extremes of an unmodulated chain (with all its couplings 
equal), so obtaining  smaller transfer times for uncontrolled state transfer 
could make preferable the task of engineering the couplings strength rather 
than implementing the control fields \cite{Nikolopoulos2015}.

Since the global optimization procedure always find a good maximum for the 
function that is being optimized, in our case the transferred population $P(T)$, where $T$ is the arrival (or transfer) time, we proceed to look for the maximum transferred population inside a hypercube
$0\leq J_i\leq J_{max}$ containing the possible values for the exchange 
couplings. 
The hypercube is the high-dimensional phase space (a 
N-dimensional hypercube of size $J_{max}$) that was mentioned in the 
Introduction. In the context of optimization methods the function that is maximized or minimized is usually called {\em cost function}.

\begin{figure}[hbt]
\includegraphics[width=0.95\linewidth]{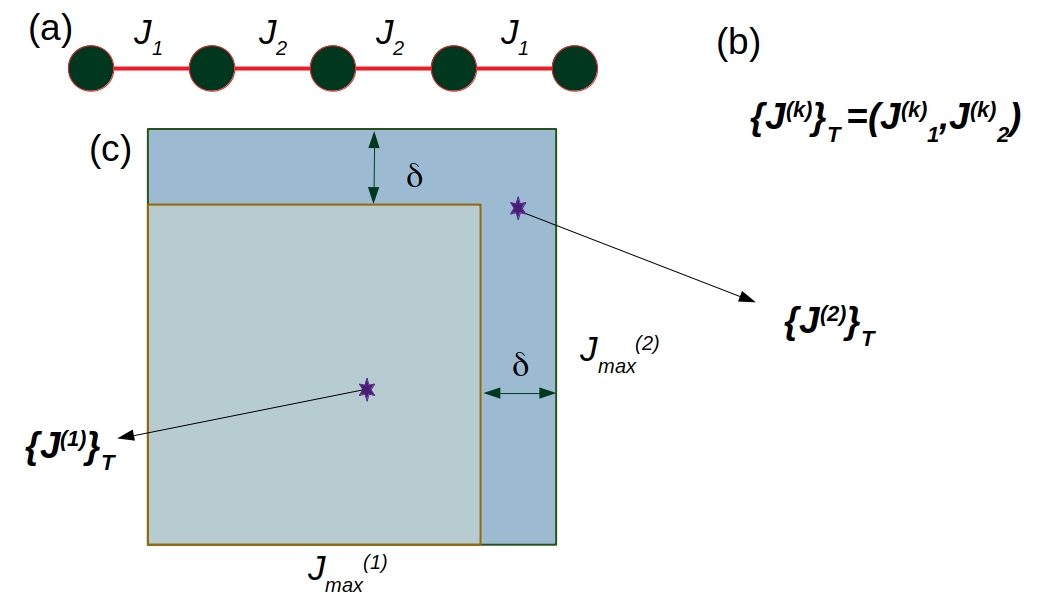}
\caption{The cartoon depicts the ingredients of the procedure designed to find EC distributions that can be used to achieve near perfect QST. a) A centro-symmetric spin chain with five spins and only two different exchange couplings $J_1$ and $J_2$. b) For the five spin chain a ECD that optimize the cost function for an arrival time $T$ is obtained using  the pivot method . The ECD is denoted by $\lbrace J^{(k)} \rbrace_T =(J^{(k)}_1,J^{(k)}_2)$, which is a point in the hypercube with side length  $J_{max}^{(k)}$. c) The smaller hypercube with side length $J_{max}^{(1)}$ and the second one smaller, with side length $J_{max}^{(2)} = J_{max}^{(1)}+\delta$, are depicted using light blue and blue squares. The corresponding ECD that optimize the cost function for each hypercube is depicted as a starry dot inside the hypercube.
}\label{cartoon}
\end{figure}

There is not a known relationship between the maximum value necessary for the couplings $J_i$ to achieve near perfect quantum state transfer at a given time $T$, for this reason
 it is useful to consider a succession of values $J^{(k)}_{max}$. For each one of this values of the hypercube side length the pivot method determines an ECD  denoted as $\lbrace J_k \rbrace_T$ where $T$ is the transfer time. The succession of  lengths of the hypercube side is given by
$J_{max}^{(k)}=J_{max}^{(1)}+(k-1) \delta J$, for $k=1,2,3,\ldots$ and $\delta J=0.5$.
As the
rigorous results about the existence of the pretty good transfer scenario are 
proved for centro-symmetric ECDs, we impose this restriction to the exchange 
couplings,  $J_{N-i}=J_{i}\;;\;i=1,\cdots,[N/2]$. The cartoon in Fig.~\ref{cartoon} depicts the main ingredients of the procedure to find the exchange coupling distributions that will lead to a chain that can attain the near perfect QST. On the other hand, it is clear that the optimization of the transferred population in a particular hypercube with side length $J_{max}^{(k)}$ provides a optimized value for the transferred population,that we denote as $P^{(k)}$, and that depends on the side length.

\begin{figure}[hbt]
\includegraphics[width=0.8\linewidth]{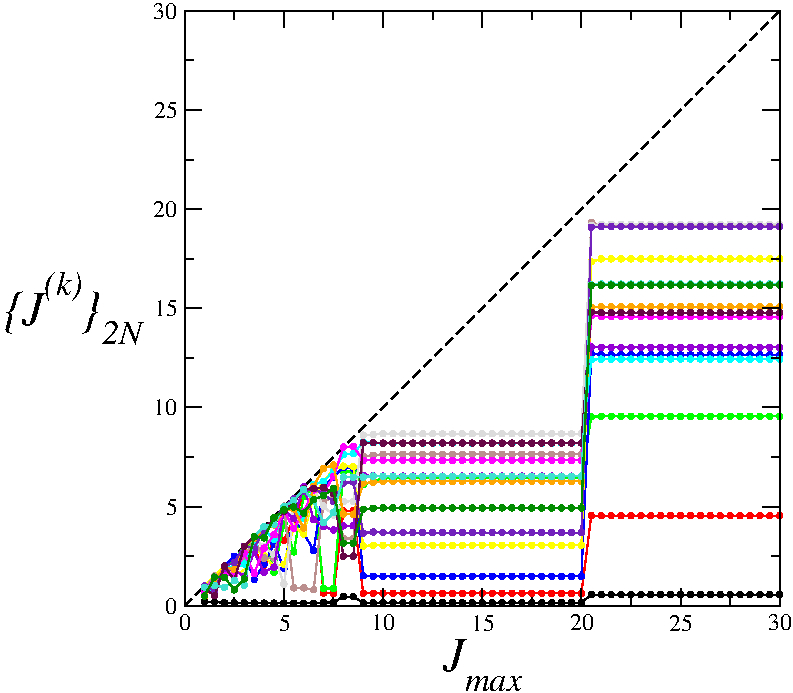}
\caption{The exchange coupling  distribution $\lbrace J^{(k)}\rbrace_T$ {\em vs} the length of the hypercube side $J_{max}$ for a system of $N=30$ spins. As explained in the text, the ECDs corresponding to a system with $N=30$ have $15$ different exchange couplings, which are shown in the figure using colored lines and symbols.  The ECDs were obtained using the pivot method for  arrival time $T=2N$ . Each set of points of a
given colour corresponds to a the values of a given exchange coupling, for 
instance the black dots correspond to $J_1$, the red ones to $J_2$ and so on. The figure clearly shows that the particular values of each exchange coupling strength $J_i$ depend on $J_{max}$. The black dashed line correspond to $J=J_{max}$ and is used a guide to the eye. Note that for small values of $J_{max}$ the exchange couplings strengths are all close to $J_{max}$, while for large values of $J_{max}$ the pivot method does not find values of $J_j$ that further optimize the population transfer and each exchange coupling $J_i$  reaches a plateau. 
}\label{fig1}
\end{figure}

Before proceeding with the analysis of the results that can be obtained using the pivot method,
 we want stress that there is not a simple way to choose the
actual value of $\delta 
J$. From 
what is known for the Heisenberg and XX models it is a fair assumption that, 
for models 
within the $XX$ universality, the pretty good transfer scenario could be 
reached with exchange couplings bounded by $N/2$, since this is the value that 
correspond to the maximal exchange coupling for the perfect state transfer case 
in the XX model. Nevertheless, the perfect state transfer happens for 
transfer times that are independent of the chain length so it can be also
expected 
that the bound on the exchange couplings are a decreasing function of the 
transferral time (this is confirmed by the results that are shown latter in the 
paper). So $\delta J=0.5$ is a 
compromise between the different scales that we thought could be present when 
dealing with models with different values of $\Delta$ and arrival times. In 
this sense, the value chosen is 
not optimal, but allow us to search for optimal values of the exchange 
coupling distribution increasing the volume of the hypercube and starting the 
search from a ``good'' initial ECD.

The workings of the pivot method are as follows. Starting
from an initial set of points in the phase space on which the cost function depends, the method explores this  space  until an 
optimal value is found \cite{ssk97,sskb97}.
In our case each point is
 an ECD that determines a Hamiltonian. Using the spectrum and eigenvalues
of the Hamiltonian the
transferred population $P(t)$ is determined for a transfer time $T$ using Eq.~\eqref{eq:population}. The pivot method
finds the largest possible value of the transferred population,
$P^{(k)}(T)$,  inside an  hypercube of side $J^{(k)}_{max}$.

Fig.~\ref{fig1} shows the typical results that can be found from a run of the procedure described in the paragraphs above. The Hamiltonian considered is the isotropic one. The figure shows the values of the fifteen different EC values in $\lbrace J^{(k)}\rbrace_T$, for a
chain with $N=30$ spins as functions of the length of the hypercube side and for a
transfer time $T=2N$. The value of
a particular coupling changes in a random-walk way when the length of the
hypercube size is
increased, eventually reaching a large plateau.

The data shown in Fig.~\ref{fig1} was obtained starting from a ECD 
with  $J_i^{(1)}=1, \forall i$. We run the  PM
algorithm  to find  the largest possible value for 
the transferred population  $P^{(2)}$ corresponding to an ECD such that  
$J_i^{(2)}\leq J_{max}^{(2)}$. Then, the length of the hypercube side was 
increased to $J_{max}=J_{max}^{(3)}$ and the PM was run again.
The 
algorithm was employed up to $J_{max}^{(K)} =N$ for some $K$. 
As usual  for stochastic global optimization algorithms, we apply to the output set a deterministic gradient algorithm \cite{numericailrecipes} in order to 
obtain a better accuracy.
Throughout this 
work, mainly in the figures, we will refer to a succession of points 
$(J_{max}^{(k)},P^{(k)}(T)) )$ as {\em the  transferred population as a 
function 
of $J_{max}$}, as if it were effectively a function. This is done for brevity 
and under the understanding that each point of the succession $P^{(k)}(T)$ must 
be calculated for an arrival time $T$  and with a specifically designed 
distribution of coefficients $J_{max}^{(k)}$. In the same sense, once an ECD 
has been designed to achieve a high  TP at a given arrival time $T$ the 
function in Eq.~\eqref{eq:population} can be evaluated at any time.

The procedure described in the paragraph above is not the most general but, as 
we will show latter, results in ECDs with very good properties, in particular 
the procedure is able to find the regime where $P(t) = 1-\epsilon$.

\section{Results: Quantum state transfer: the Heisenberg
Hamiltonian}\label{sec:Heisenberg}

\begin{figure}[hbt]
\includegraphics[width=0.8\linewidth]{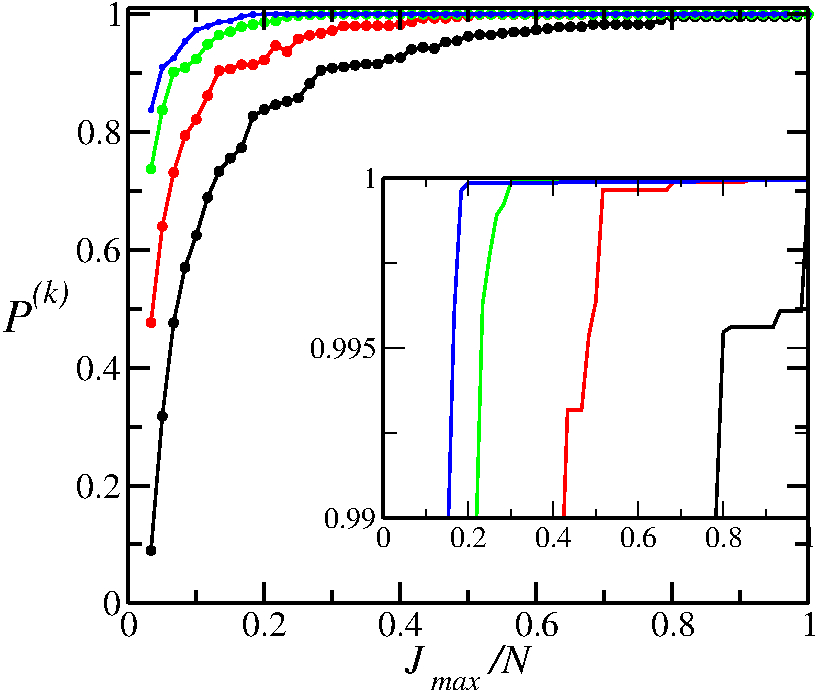}
\caption{The transferred population for a chain with Heisenberg Hamiltonian and 
$N=30$ spins. The data corresponds to arrival times $T=N/2, N, 2N$ and $3N$, 
which are depicted as black, red, green and blue dots, the lines joining the 
dots are used as a guide for the eye. Each point corresponds to a pair 
$(J_{max}^{(k)}/N, P^{(k)})$, look at the text for definitions. The inset shows
a zoom of the region where the transferred population exceeds the value 
$0.99$.
}\label{fig2}
\end{figure}

The corresponding values of the 
transferred population $P^{(k)}(T)$ obtained with the ECD shown in 
Fig.~\ref{fig1} are shown in Fig.~\ref{fig2}, which  also shows 
transferred population  corresponding to arrival times $N/2$ and $N$ for a 
chain 
length $N=30$. Each point corresponds to a pair $(J_{max}^{(k)}/N,P^{(k)}(T))$, 
the lines are included as a guide for the eye. From the data sets it is clear 
that in general the ordering $P^{(k)}(T_<) < P^{(k)}(T_>)$ holds, where $T_> > 
T_<$ are two ordered arrival times. However, since $P^{(k)}(T)$ is a discrete 
sample of a random-like quantity for some $k$'s the ordering is reversed. All in 
all, the data in the figure is compatible with two reasonable physical traits, 
i) for a fixed $J_{max}^{(k)}$  larger arrival times results in larger 
transferred population, and ii) to reach a given target transferred population  taking larger 
arrival times 
ensures that the target is attained for lower  $J_{max}^{(k)}$.

The inset in Fig.~\ref{fig2} shows a detailed view of the region where 
the 
transferred population  $P>0.99$. This zoom level allows us to appreciate that 
the transfer is of a very high quality and this is due to the excellent 
properties of the coefficient distribution found by the optimization method. 
Another feature that can be clearly observed in the inset has to do with the 
decreasing value of $J_{max}$ for which the transferred population reaches a 
given preset value, say $P=0.99$. The inset shows that this value is reached 
for $J_{max} \sim 0.8 N$ for an arrival time equal to $N / 2$, $J_{max} \sim 
0.4 N$ for an arrival time equal to $N$, etc. Even at this 
level the of magnification the largest values of $P^{(k)}$ attained for the 
largest values of $J_{max}^{(k)}$ explored can not be fully appreciated, since 
they are larger than $0.999$.

The results presented so far in the figures correspond to a chain of 
length $N=30$, in any case we have checked that the algorithm to find all the 
coefficients of a chain works correctly for chains with lengths between 10 and 
90 and for various arrival times, mainly $N /2, N, 2N, 3N, 4N$  and $5N$. In 
the Supplemental Material \cite{sup-mat} we include a table with the 25 
different coefficients of a chain of length $N=50$, for the times mentioned, in 
addition to the value of the transferred population that is obtained when a 
chain is assembled with these exchange coefficients. The coefficients in the 
Table will be used latter on.

From a theoretical point of view, that the regime where $P=1-\epsilon$ can only 
be attained using $J_{max}^{(k)} \approx N$ is not an impediment or a 
problem. For an actual implementation this trait could render the ECDs 
responsible of that regime useless, since  changing the 
strength of the interaction  between the components of a microscopic system on 
one or two orders of magnitude (depending on the length of the chain) is 
extremely difficult. So, in the rest of this Section we will focus on the 
properties of the ECDs obtained with small or moderate values of 
$J_{max}^{(k)}$ to evaluate them as alternative distributions that can be 
experimentally implemented.

One concern about the ECDs obtained for small values of $J_{max}$ is that the 
transferred population  obtained from them could be unstable under static 
perturbations of the couplings, so the transferred population  would be easily 
ruined regardless of the very good quality of the transfer produced by the 
unperturbed distribution.

\begin{figure}[hbt]
\includegraphics[width=0.8\linewidth]{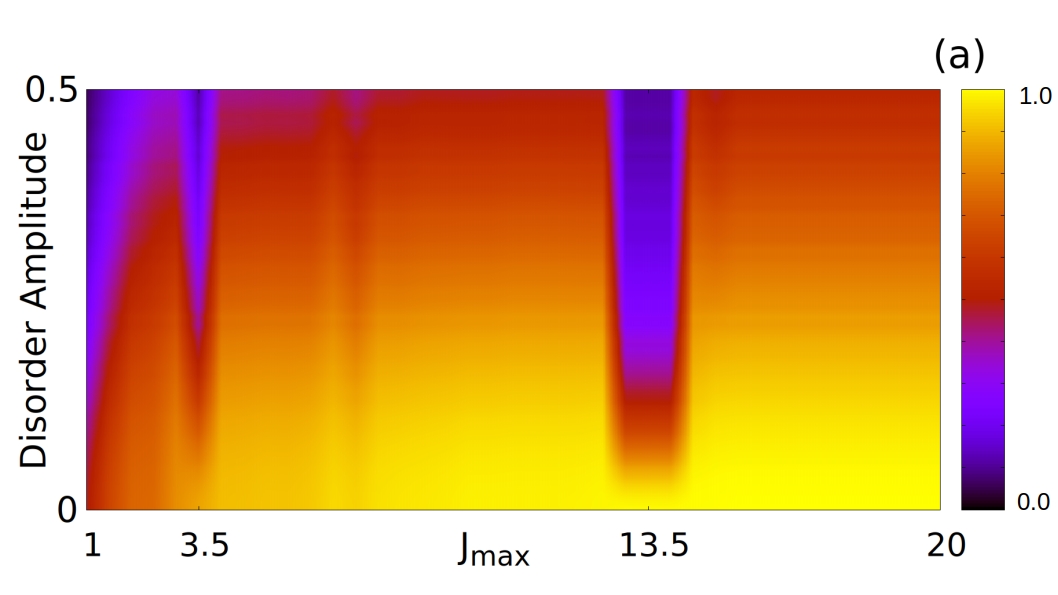}
\includegraphics[width=0.8\linewidth]{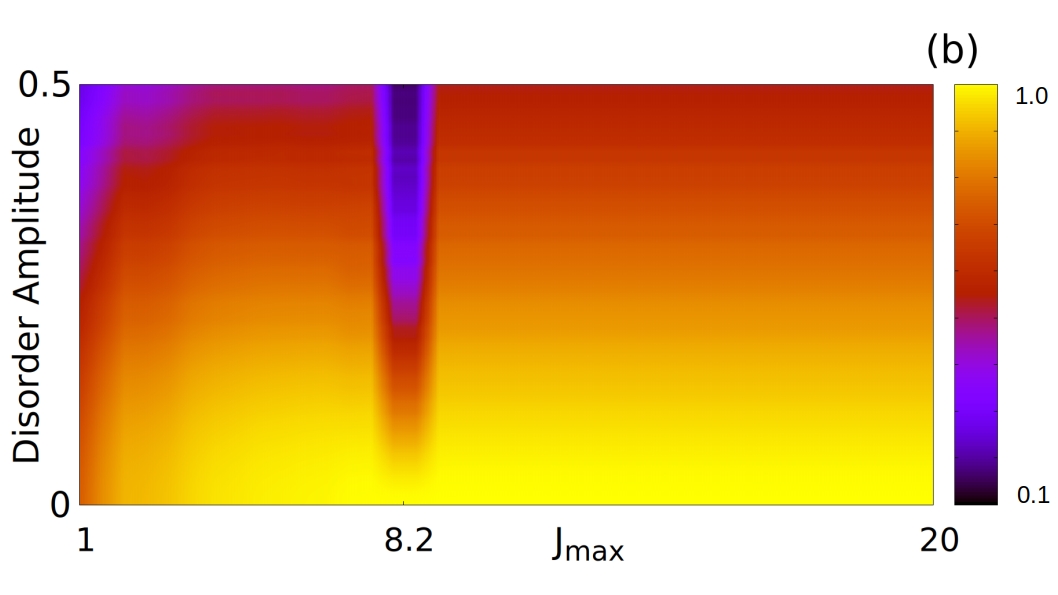}
\includegraphics[width=0.8\linewidth]{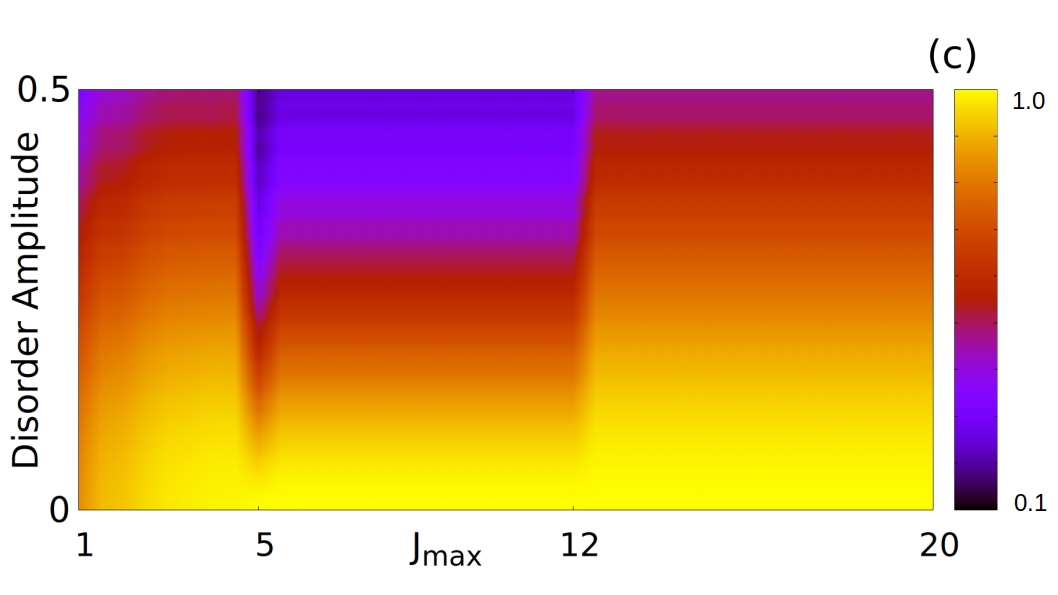}
\caption{Color map of the averaged transferred population as a function of the
maximum value of the exchange coupling $J_{max}$ and the  
static disorder amplitude. Calculations were performed for a chain with
$N=30$ spins and for different arrival times: (a) $T=N$, (b) $T=2N$
and (c) $T=3N$
}\label{fig3}
\end{figure}

The stability of quantum state transfer in spin chains under static 
perturbation, or static disorder, {\em i.e.} the robustness of the  
transferred population when the  ECDs is subject to static perturbations has 
been 
extensively studied in the literature. The static disorder is understood as a 
consequence of  manufacturing errors  of the physical system that is modeled by 
the spin chain. In theoretical studies the static perturbations are modeled 
adding a random term proportional to the coupling strength 
\cite{Zwick2,Zwick3,Petrosyan2010}

\begin{equation}
\label{eq:couplin-disorder}
J_i \longrightarrow J_i (1 + a \xi_i) ,
\end{equation}

\noindent where $a$ is the strength of the disorder and $\xi$ is a random 
variable with zero mean and unity variance. 

When static disorder is added, the transferred population averaged over 
realizations of the disorder is introduced as a figure of merit

\begin{equation}
 \bar{P} = \left\langle P \right\rangle_{\xi} ,
\end{equation}

\noindent where the average is a simple equally weighted average. This means 
that for a chain with $N$ spins a realization of the disorder implies that 
$N-1$ random numbers are generated $\lbrace\xi_i^j\rbrace$ and the ECD to be 
analyzed is changed as
\begin{equation}\label{eq:a-realization-of-J}
 J_i \longrightarrow J_i (1+a \xi_i^j) ,
\end{equation}

\noindent where $1\leq j \leq n_r$, and $n_r$ is the number of realizations to 
be taken into account. Once a particular ECD has been generated, 
Eq.~\eqref{eq:a-realization-of-J}, the corresponding transferred population at 
arrival time $T$ is calculated, $P^j_{\xi}(T)$, and the averaged TP is 
calculated as
\begin{equation}
 \bar{P}(T) = \frac{1}{n_r} \sum_{j=1}^{n_r} P^j_{\xi}(T).
\end{equation}

Fig.~\ref{fig3} shows, as color maps, the averaged transferred population 
(TP)  as a function 
of $J_{ax}^{(k)}$ and the disorder strength for a chain with $N=30$ and three 
different arrival times. As the color map indicates, there are segments of the 
$J$-axis where the TP is particularly unstable, so even for 
very small disorder strengths the TP becomes quite poor 
($P<0.9$), note the dark orange and blue fringes that appear in every map shown. 
Nevertheless, the segments become sparser and the fringes appear for larger 
disorder strengths when the arrival time considered is longer, and, for the 
three cases depicted there are always regions where the TP is 
very high and robust (the brilliant yellow regions) even for small $J$-values.

For further analyze the properties of ECDs found using the PM that produce 
fairly good state transfer ($P\gtrsim 0.95$) it is useful to look back at the 
time evolution of the population transfer, Eq.~\eqref{eq:population}. So far we have studied the values reached by the population transfer at the times for which the ECD was optimized. Now we want to study the population transfer as a continuous function of time, Eq.~\eqref{eq:population}. It is clear that the time-dependent functions will be different depending on the ECD that is considered but, as we will show, these different functions are closely related.

In this paper we consider that $T = \kappa N$, where $\kappa=1/2,1,2,3,\ldots$. It is clear that the population transfer obtained with an ECD optimized to show pretty good transmission at $T$ will show successive peaks at times $T_p = (2p+1) T$, where $p=1,2,3,\ldots$,  since the time evolution should be almost periodic and the time for the state to go back and forth between the extremes of the chain is $2T$. For perfect state transfer the successive peaks of the TP are all of the same height, but for pretty good quantum state transfer the height will not be constant. The TP then, as a function of the scaled time  $t/N$, will show 
peaks at the scaled arrival times $T^{sc} = (2p+1) \kappa$.

\begin{figure}[hbt]
\includegraphics[width=0.8\linewidth]{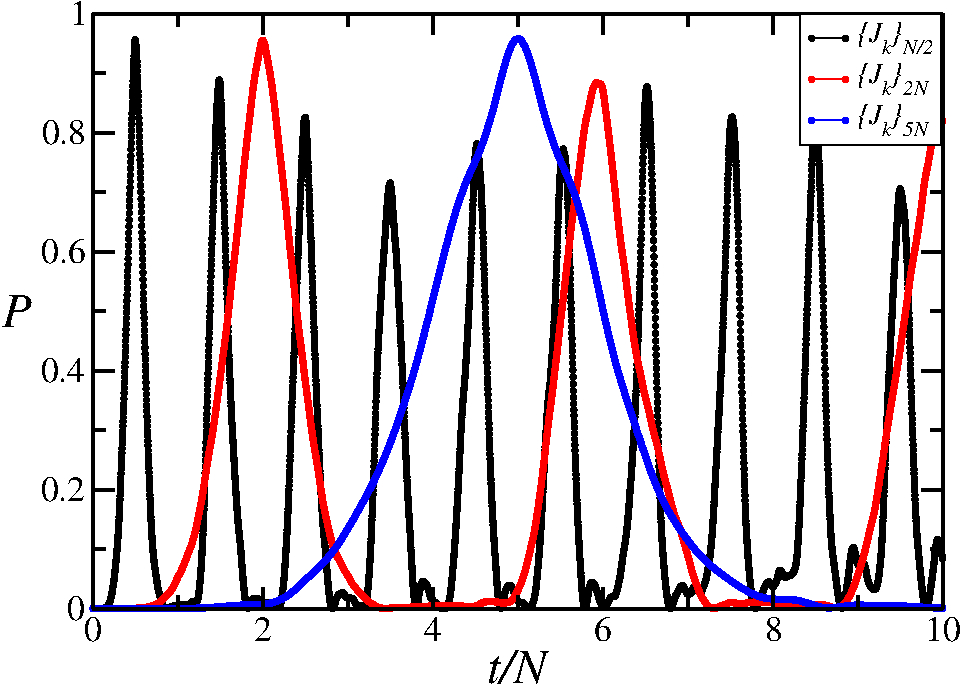}
\caption{The transferred population obtained with a Heisenberg chain with 
$N=50$ {\em vs} the scaled time $t/N$. The TP corresponding to three 
different arrival times, $T = \kappa N$, with $\kappa=1/2, 2$ and $5$ is shown using black, red and 
blue curves, respectively. Note, for instance, that for $\kappa=1/2$ the black curve shows peaks at $t/N=1/2,3/2, 5/2, \ldots$ The ECD for each case was designed to achieve a target population 
of $0.95$. The particular values of the coupling coefficients  used to assemble 
the transferring  chains can be found in the Supplemental Material 
\cite{sup-mat}.
}\label{fig4}
\end{figure}

\begin{figure}[hbt]
\includegraphics[width=0.8\linewidth]{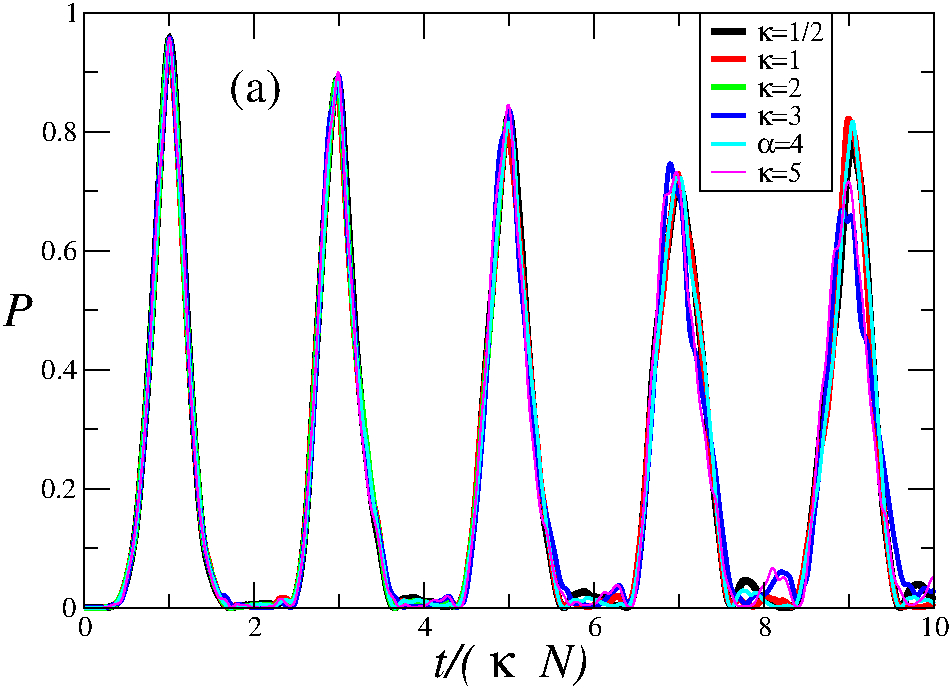}
\includegraphics[width=0.8\linewidth]{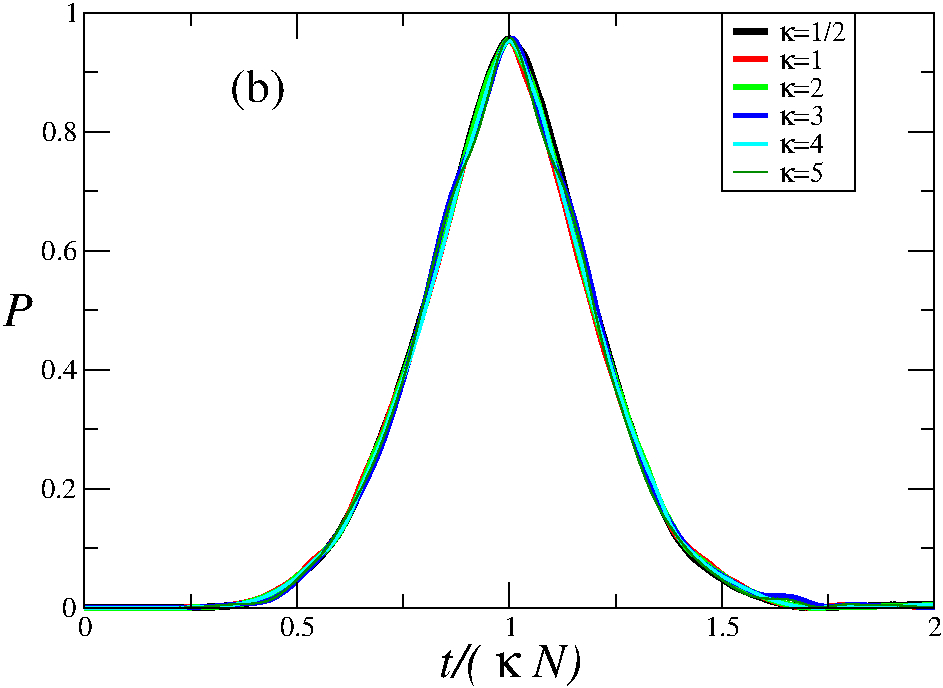}
\caption{a) The first five peaks of $ P(\lbrace J_k\rbrace_{\kappa N},t)$ {\em vs} 
the scaled time $t/(\kappa N)$ for $\kappa=1/2, 1 , 2, 3, 4$ and $5$, are shown using 
purple, magenta, black, red, green and blue curves, respectively.. The excellent 
superposition between the curves does not allow to distinguish between them if 
several peaks are included, because of this a single peak is shown in b). The 
particular values of the coupling coefficients  used to assemble 
the transferring  chains can be found in the Supplemental Material 
\cite{sup-mat}.
}\label{fig5}
\end{figure}

Fig.~\ref{fig4} shows the time evolution of the TP  for 
three different arrival times. The corresponding ECDs were designed using the 
PM algorithm. The TP calculated using the ECD that is designed to 
produce a fairly good TP at arrival time $N/2$, the black curve,  shows several peaks 
at scaled times $t/N= 1/2, 3/2, \ldots, (2j+1)/2, \ldots$, for $j\in 
\mathbb{N}$. The TP calculated using the ECD that must produced a peak at arrival time 
$2N$ shows several peaks at scaled times $2, 6, 10, \ldots$, while the 
horizontal scale only allows for a single peak of the TP calculated using the 
ECD that corresponds to $T= 5N$. It can be seen that the peaks are broader for 
larger arrival times. However the apparent differences in time scale and width 
of the peaks corresponding to different arrival times, changing the scaling on 
the horizontal axis reveals a more profound relationship between them, as shown 
in Fig.~\ref{fig5}.

Calling $P(\lbrace J_k\rbrace_T,t)$ the TP calculated using an ECD optimized for an arrival 
time $T$, Fig.~\ref{fig5} shows the first five peaks of $P(\lbrace J_k\rbrace_T,t) $ {\em vs} the scaled time $t/(\kappa N)$, for $\kappa =1/2,1,2,3,4$ and $5$. The 
scaling is the cause the curves superposition and reveals the physics 
involved in the TP. 

Although the ECD for each chain size and arrival time is obtained using a 
stochastic method, it is interesting to note that the distributions thus 
obtained have common features that allow deciding, according to the strength of 
the interactions available in a particular implementation, which ones are the 
arrival times compatible with these strengths.

So far, our results show that it is possible to find ECDs such 
that near perfect quantum state transfer is achieved for arrival times as 
short as 
$T=N/2$, if the strength of the interaction can be taken large enough. This is 
not what happens in actual implementations, where the range of strengths at 
disposal are limited by technological or physical reasons. So, it is 
reasonable to ask how to proceed when the range of interactions is limited. The 
answer, in a nutshell, is that the ECD should be sought at longer arrival 
times. In what follows we will show that the ECDs found using the pivot method 
show a relatively simple behavior that helps to determine the appropriate 
arrival time compatible with the range of available interactions.

We look for the smallest hypercube that contains values of the 
exchange coefficients such that the transferred population reaches the target 
value of 0.98 and we call the corresponding side length value $J_ {min}$, which 
depends on the length of the chain and the arrival time.

\begin{figure}[hbt]
\includegraphics[width=0.8\linewidth]{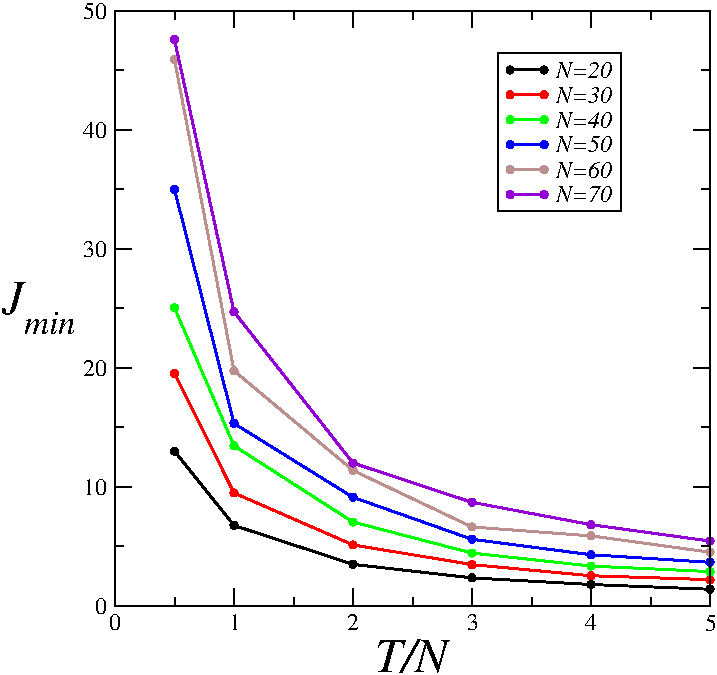}
\caption{The minimum length of the hypercube side necessary to obtain a 
population transfer of 0.98, $J_{min}$,  versus the transfer time, for chains 
of length (from bottom to top) $N = 20, 30, 40, 50, 60$ and $70$, the data is 
shown using  black, red, green, blue, etc, solid dots, respectively. The 
lines are included as a guide for the eye  }.
\label{fignew}
\end{figure}

Fig.~\ref{fignew}  shows  
$J_{min}$ for 
different chain lengths and arrival times. The values can be fitted, with a 
very good 
agreement,  using an exponential function 

\begin{equation}\label{eq:exponential-fit}
 J_{fit} = A(N) + B(N) e^{-C(N) \frac{T}{N}} .
\end{equation}

The data and the fitting show that for fixed target TP the 
side length is a rapidly decreasing function of the arrival time. We 
think that this 
type of behavior is rather general if the target population is low enough 
(0.98, or 
lower) and the chain is long enough. The constants $A$, $B$ and $C$ on 
Eq.~\eqref{eq:exponential-fit} depend on  the chain length, although $C(N)$ seems 
to be independent of $N$ once the uncertainties in the data are taken into 
account.

Although the results shown in figure 6 tell us that for long arrival times the intensities of the exchange interactions must all be of the same order, it is interesting to note that there are different actual physical systems in which it is possible to tune the particular values of said interactions on ranges and signs compatible with those studied in the present work. For instance, in Ref.~\cite{Kandel2021} which deals with the perspectives of exchange coupled quantum dot spin chains, it is shown that positive exchange interactions can be tuned between 0 and 125 Mhz. The tunability can be used to implement SWAP operations between qubits located in adjacent quantum dots or to fix each exchange interaction coupling to a desired value. In Ref.~\cite{Martins2017} it is shown that the sign of the exchange interactions can be chosen and it can be tuned between 0 and 40 $\mu$eV. It is worth noting here that the one-excitation transfer protocol works in the same way for spin chains with all the exchange interactions positive or negative. The tunability of the exchange couplings strengths also has been demonstrated in chains of superconducting qubits see, for instance \cite{Li2018}.

Before changing the model, it is interesting to note some of the 
consequences of our results with respect to the ``pretty good quantum state 
transfer" scenario, since until now
the results presented in this Section do not include  the chain lengths where 
the regime $P=1-\epsilon$ can be proven to exist. 

 As well as the results shown explicitly in figures , we also 
obtained results for chains with  even and odd number of spins, in addition to 
chains with length equal to powers of two . We have explored the behavior of 
chains with lengths between $N=10$ and $N=90$ spins, while comparing 
the results of chains of lengths $N=30$ and $N=31$ with $N=32$  we have not 
found qualitative differences between the largest transfer values accessible for any pair of chains that were compared. We  also compared 
the results for chains with   $N=60$ and $N=61$ with $N=64$ and, again, there
were not discernible qualitative differences. This study was also made for 
different 
transmission times, which once again, showed that there are no qualitative 
differences in the largest values that can be obtained for the population 
transferred in chains of similar length, beyond their length being even, odd or 
a power of two. We think this presents fairly conclusive 
evidence that such a scenario is accessible for spin chains of any length with 
the Heisenberg Hamiltonian

It is interesting to wonder if the coupling distributions found 
using the pivot method allow a robust transmission of states that are initially 
not pure due to problems with their preparation. The Heisenberg Hamiltonian 
commutes with the total magnetization for any set of values that the ECDs adopt, 
and this can be used to estimate the population transfer in the case that the 
initial state is an ``almost pure" state such as those used. in Nuclear Magnetic 
Resonance experiments.

Let us suppose that the state prepared initially can be written as
\begin{equation}
 \rho_i = (1- \eta) |\psi \rangle \langle \psi|\otimes |\mathbf{0}\rangle 
\langle \mathbf{0}| + \eta \rho_u ,
\end{equation}
where $\psi$ is 
the pure state that is intended to be prepared at the first site of the chain 
and  transmitted with
\begin{equation}
 1-\eta = \langle \psi| \otimes \langle \mathbf{0}| \rho_i |\psi \rangle 
\otimes |\mathbf{0}\rangle 
\end{equation}
so $\rho_u$ accounts for all the undesirable parts of the 
state that can not be eliminated because the preparation of the state is not 
``perfect'', and $\eta\ll 1$.
Then the state after the transmission protocol can be written as
\begin{equation}
\rho(t) = U(t) \left((1- \eta) |\psi \rangle \langle \psi| \otimes 
|\mathbf{0}\rangle 
\langle \mathbf{0} + \eta \rho_u 
\right) U^{\dagger},
\end{equation}
and the fidelity with the ``expected'' state is given by
\begin{equation}
F = \langle \psi| \mbox{Tr}_{N-1}(\rho(t)) |\psi \rangle .
\end{equation}

So, the transferred population of an ``almost pure'' state would be bounded by
\begin{equation}
 P_{\rho} \sim P (1-\eta),
\end{equation}
where $P$ is the transfer probability of the one-excitation protocol.
In the paper by Di Franco et al. \cite{DiFranco2008}, it is discussed how to 
obtain perfect transfer in the event that there is no ``initialization of the 
state to be transmitted", that is, a series of gates are applied to the ends of 
the chain, which allows perfect transmission  of a mixed or pure state when the 
chain has been designed for perfect transmission in the sub space of an 
excitation. Our estimate of an upper bound for the transmission of poorly 
prepared states is consistent with the results of Di Franco et al. .

\section{Results: Quantum state transfer for chains with XXZ Hamiltonian}\label{sec:XXZ-Hamiltonian}

It has been suggested by Kay \cite{Kay2019} that some transfer properties of 
quantum spin 
chains with exchange couplings roughly equal to those of the Heisenberg 
chain can be studied using perturbation theory. There are several reasons for 
doing that study. In particular, it is of great interest to study how the 
transfer properties of the Heisenberg Hamiltonian, with their inherent 
difficulties, change towards the properties of the XX Hamiltonian, more direct 
to understand and for which several transfer regimes have been identified that 
show a direct relationship between the quality and quantity of modulation 
exerted on the coefficients and the quality of the transfer obtained by said 
modulation.

\begin{figure}%[hbt]
	\includegraphics[width=0.7\linewidth]{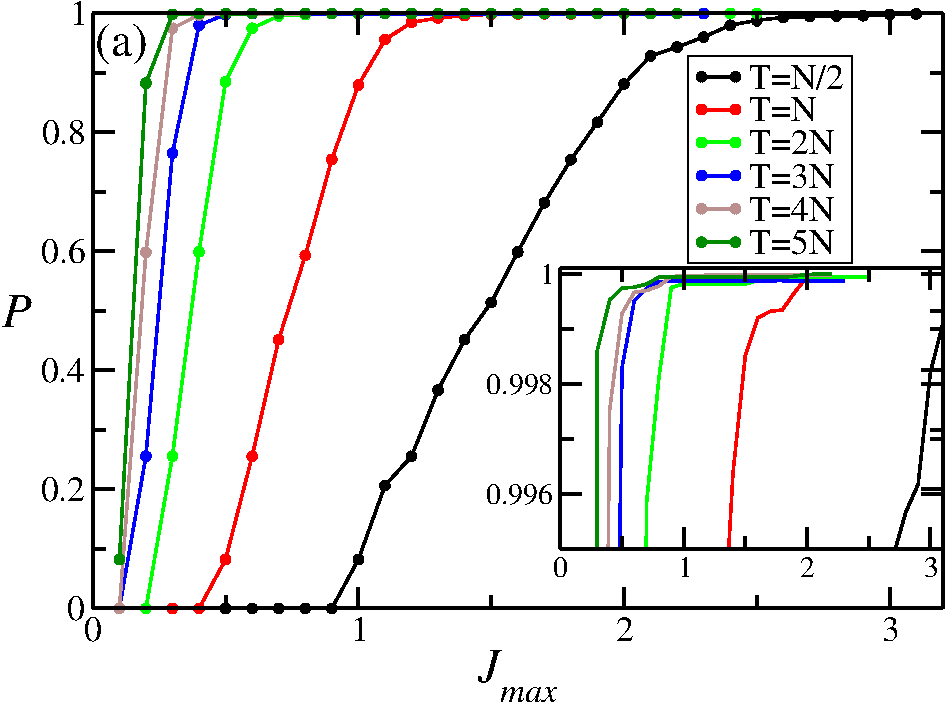}
\includegraphics[width=0.7\linewidth]{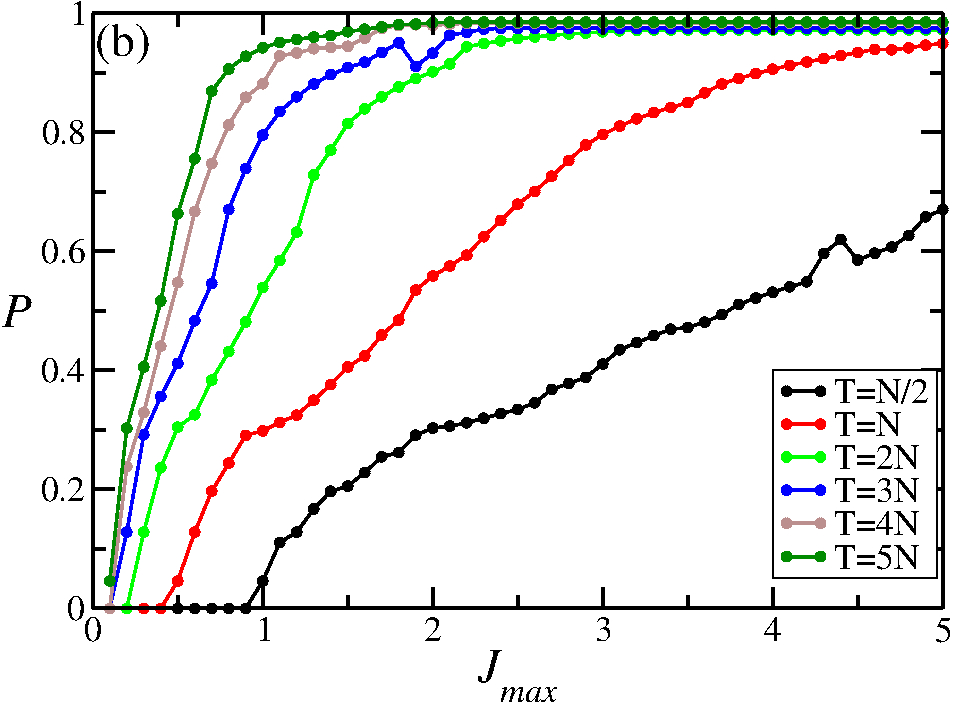}
\caption{The transferred population $P$ {\em vs} $J_{max}$ for two chains with 
$N=50$ and a) $\Delta=0.8$, b) $\Delta=1.2$. The 
data included in both top panel correspond to arrival times $N/2$, $N, 2N, 
3N, 4N$ and $5N$. The behavior shown by the transferred population in the top 
panel shows that, even for the shorter arrival times, the values of the 
exchange couplings needed to attain $P>0.999$ are bounded by $J_{max}\lesssim 
3$. In the case $\Delta=1.2$, shown in the bottom panel, the values of 
$J_{max}$ needed to attain high quality transfer, $P>0.99$ can become quite 
large for short arrival times, in a similar way to the Heisenberg case, see 
Fig.~\ref{fig2}.
}\label{fig6}
\end{figure}

In this Section we apply to anisotropic Hamiltonians with $\Delta\neq 1$, see 
Eq.~\eqref{eq:XXZ-Hamiltonian}, the same methods that we have applied to the 
Heisenberg Hamiltonian in Section~\ref{sec:Heisenberg}.

Fig.~\ref{fig6} shows the behavior of the TP as a function of 
the side length of the hypercube in whose volume the global optimization 
algorithm looks for the distribution of couplings that produces the largest 
possible transfer. The results shown in panel a) correspond to $\Delta=0.8$, 
while the results shown in panel b) correspond to $\Delta=1.2$, respectively.

Comparing the TP shown in the two panels of Fig.~\ref{fig6} it is 
clear that the 
case with $\Delta = 0.8$ transfers considerably better and with 
the difference being greater for shorter arrival times. It is important to note that in the case 
of the figure corresponding to the Heisenberg Hamiltonian case, 
Fig.~\ref{fig2}, the TP was shown as a function of the 
length of the hypercube side renormalized by the length of the chain, while in 
 Fig.~\ref{fig6} it is shown as a function of the length of the side without 
any renormalization. The inset in Fig.~\ref{fig6} a) shows  the 
TP attaining values well above 0.999 for every arrival time 
considered.

As was said in  Section~\ref{sec:Heisenberg}, when it 
comes to qualifying the performance of a chain built with a given distribution 
of coefficients, it is not only necessary to take into account the value of 
$J_{max}$ that limits the possible values of the coefficients and the magnitude 
of the transferred population, in addition  the stability of the  
TP must be studied in the face of static disturbances.

\begin{figure*}%[hbt]
\onecolumngrid

\includegraphics[width=0.95\linewidth]{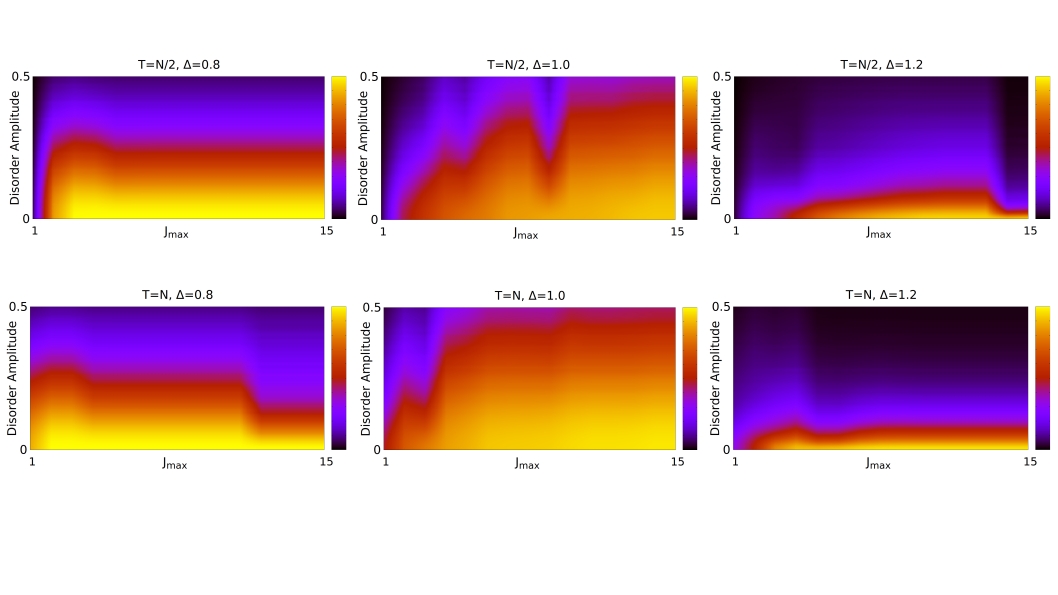}
\caption{Color map of the averaged TP as a function of the maximum
value of the exchange coupling and the static disorder amplitude.
Results were obtained for a chain with $N=51$ spins, two different
arrival times $T=N/2$ and $T=N$ and three different values of the
anisotropy parameter $\Delta=0.8, 1.0$ and $1.2$.
}\label{fig:Klee-Fe-Se}
\twocolumngrid

\end{figure*}

Fig.~\ref{fig:Klee-Fe-Se} shows the behavior of the averaged TP for three chains
with the 
same length, $N=51$. The three columns, from left to right, correspond to 
Hamiltonians with $\Delta = 0.8, 1$ and $1.2$. All the necessary ECD were 
obtained using the PM. The top row correspond to the shorter arrival time, 
$T=\left[\frac{N}{2}\right]$, while the bottom row correspond to the larger 
arrival time $T=2\left[\frac{N}{2}\right]$. The figure summarizes quite well 
our 
findings: i) For moderate or small values of the maximum value allowed to the 
exchange coefficients, the figure shows values up to $J_{max}=14$, both 
anisotropic chains attaining larger TP when the strength of the static 
disorder  is negligible, but the chain with $\Delta$ larger than unity is 
very strongly affected by disorder, making the chain useless for disorder 
strengths larger than $0.1$, as can be seen in the two panels in the right 
column and shown as a narrow fringe of brilliant yellow and light orange, the 
two colors that correspond to a high-quality transfer. ii) Conversely, the TP 
obtained using the chain with $\Delta = 0.8$ is weakly affected by static 
disorder, presenting a TP of a very high quality, as is depicted by the 
broad fringe of brilliant yellow and light orange that appear in both 
panels in the left column. iii) Finally, 
the Heisenberg chain is very resistant to static disturbances, even for a range 
of intensities of the disorder even greater than the chain with $\Delta <1$. 
However, the first type of chain does not reach as high transmission values as 
the second one for these values of $J_{max}$.

The behavior of the TP shown in the right and left columns of
Fig.~\ref{fig:Klee-Fe-Se} can be, in practice, translated to any value of 
$\Delta>1$ and $\Delta<1$, respectively. Chains whose Hamiltonians 
have $\Delta<1$ and ECDs optimized using the PM show excellent quantum state 
transmission, with very high values of $P$ and are very resilient against 
static disorder. Moreover, the ECDs obtained for short arrival times  are 
bounded by $J_{max} \sim \mathcal{O}(1)$ independently of the chain length, at 
least for $N<100$  the  range of chain lengths that we have explored. In a 
sense, it is reasonable to say that chains with Hamiltonians with $\Delta<1$ 
are XX-like in as much as their quantum state transfer properties are involved.

\begin{figure}[hbt]
\includegraphics[width=1\linewidth]{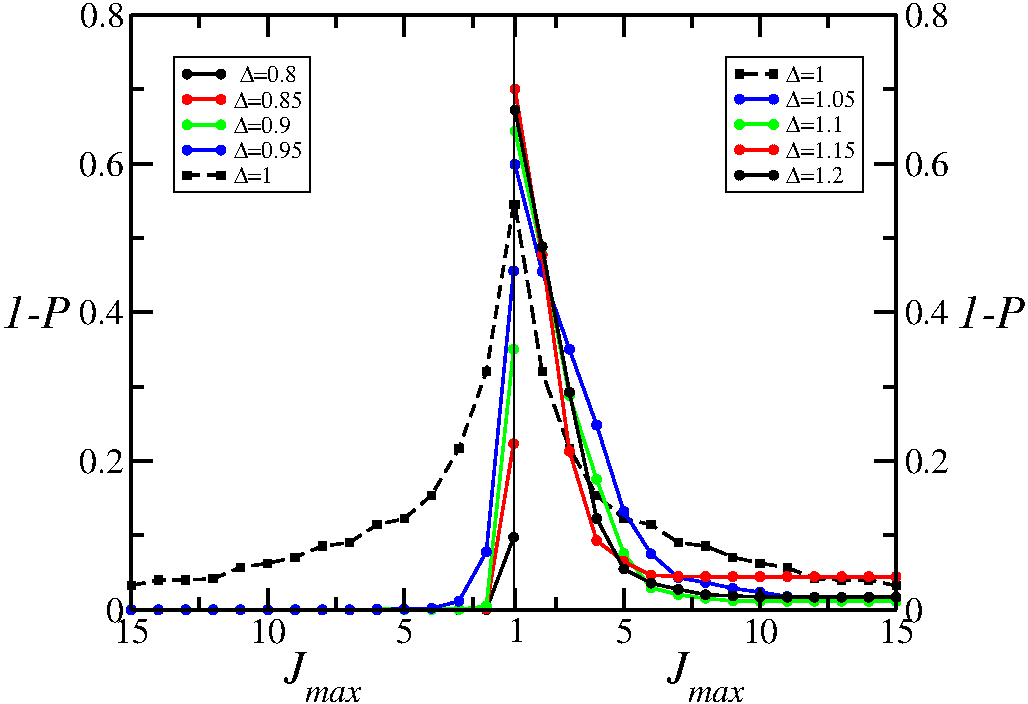}
\caption{$1-P$ {\em vs} $J_{max}$ for chains with $N=61$ and for an arrival 
time $T=60$. The transferred population is shown in this way to emphasize the 
quite different behavior observed by the transmission in chains with $\Delta < 
1$,whose data is shown in left half of the figure, from the one observed in 
chains with $\Delta>1$, whose data is shown in the right half of the figure. Note that the values of the variable $J_{max}$ increase from 1 (in the middle of the  horizontal axis) to 15 (at the left and right extremes of the horizontal axis)  . The same colors 
are used for data that correspond to chains with the same  $\Delta^{\prime}$, 
with $\Delta=1\pm \Delta^{\prime}$. Note that $\Delta^{\prime}=0.2$ is shown 
with black dots,  $\Delta^{\prime}=0.15$ with red dots and so on. 
}\label{fig8}
\end{figure}

 The rapid departure of the behavior of the transferred population when the 
Hamiltonian of the chain has $\Delta<1$, from the behavior 
observed when the chain obeys the Heisenberg Hamiltonian can be seen in the 
Fig.~\ref{fig8}. The figure shows $1-P$ as a function of $J_{max}$, in the 
left panel the data corresponds to chains with $\Delta\leq 1$, while the right 
panels shows the data corresponding to chains with $\Delta\geq 1$. The color of 
the data points and curves in both panels have been chosen so that the chains 
at the same "distance" from the isotropic case have the same color, so
the 
data for the chains with  $\Delta=0.8$   and 1.2 are shown with black points, for the
chains with   $\Delta=0.85$ and 1.15 with red points, and the following with green
and blue dots. The resemblance with `` $\lambda$-point'' plot is striking.

The value of 
$J_{max}$ for 
which the PT reaches a given target value for chains with $\Delta>1$ is an 
increasing function of $1/(1-\Delta)$. To determine this function, it would be necessary to study much 
longer chains, with the consequent expenditure in computing time, anyway for 
finite chains there can not be any divergences so, in the worst case scenario 
$J_{max}$ grows up to values $\mathcal{O}(N)$  for arrival times 
$\mathcal{O}(N)$.

Even when the averaged transferred population could be  very high, it is logical to ask what the "worst case" scenario is like. This question is particularly important in experiments in which only a single-shot transmission  attempt is made and there are no error correction algorithms available. To obtain the worst case scenario, it is necessary to calculate numerous realizations of the disorder and keep track of the smallest transferred population.

In the Supplemental Materials we have included a figure comparing the averaged transferred population and the worst transferred population as functions of the disorder amplitude and  for different chains . We also included a plot showing the standard deviation of the transferred population. All in all, the results, although preliminary, show that for moderate disorder amplitude the Heisenberg isotropic chain is the better option as a transmission channel and that the statistics of transferred population in quite different between the different model chains. Other authors also have studied the statistics of the quantum state transfer, in particular G. Nikolopoulos \cite{Nikolopoulos2013}. In this paper it is shown that the minimal fidelity is equal to the worst population transferred for $P>0.5$. The regime of interest is found for values of $P>0.5$, so the worst case TP is a good qualifier of the transfer process.

\section{Discussion and Conclusions}\label{sec:conclusions}

If the strength of the couplings is not an issue, which in a microscopic system 
does not seem realistic, it is possible to design chains with the Heisenberg 
Hamiltonian that transfer an excitation between the ends of the chain with very 
good probability at times as short as $N / 2$ or $N$. In any case, if the 
intensity 
of the couplings is limited (at most couplings less than or equal to two), 
considering arrival times of the order of 4 or $5N$ is recommended. This last 
point is of particular importance for possible practical implementations. In 
the same sense, if the spin Hamiltonian is an effective approximation for the 
low energy spectrum of a given physical system, it is advisable to look for a 
physical regime for the original system in which the resulting model 
Hamiltonian  is anisotropic with $\Delta<1$.

Based on the results shown Fig.~\ref{fig:Klee-Fe-Se} about the behavior of 
the averaged TP we analyzed the stability  of the 
transmission process for relatively short arrival times, $N / 2$ and $N$. We 
also obtained results for longer times and the same chain length $N=50$, which 
are not consigned to any figure, and the conclusion drawn in 
Section~\ref{sec:XXZ-Hamiltonian} holds, that is, chains with $\Delta<1$ 
transfer excitations with greater probability than chains with $\Delta=1$ , 
which in turn transmits better than those with $\Delta>1$. For  times $4N$ or 
$5N$, the chains with $\Delta \leq 1$  reach very high values of the 
transmission probability ($P >0.9998$), but in  those with Heisenberg's 
Hamiltonian the averaged TP decay faster than on the chains with $\Delta<1$ when 
the intensity of the disorder is increased . The chains with $\Delta>1$ never 
reach the highest levels of quality in transmission than are attained by chains 
with $\Delta \leq 1$ and are severely affected by static disorder.

The linear scaling of the transferred population, Fig.~\ref{fig5}, is 
noteworthy for a number or reasons, i) the ECD which are used to obtain the 
PT are bounded by a quantity that changes non-linearly with the arrival time, 
ii) each ECD is obtained independently for each time, and iii) the TP is a 
nonlinear functions of the couplings. So far we have not reached a conclusion 
about how to look for scaling properties in the eigenvalues of the Hamiltonian 
of the chain and the coefficients of the eigenvectors that enter into the 
calculation of the transferred population.

The results found show us that the ECDs found are relatively robust but, even 
in the best case, if there is a static disorder present with a strength 
larger than $10\%$, these distributions no longer offer a valid alternative to 
build chains of spins that serve for the transfer of quantum states. In the 
case of chains with XXZ Hamiltonians with $\Delta> 1$ even smaller disorder 
strengths render useless the chain as a transmission channel. Some of these 
features can be better appreciated in the figure that is included in 
\cite{sup-mat}. So, from the 
quantum state transfer point of view the regime of interest can be found for 
$a< 0.1$. Notwithstanding this, the patterns observed in Fig.~\ref{fig3} 
and \ref{fig6}, which  can be clearly appreciated  for disorder strengths 
larger than $0.1$, seem to indicate that there are different dynamical regimes. 
The propagation of information can show very different speed 
bounds in disordered spin chains and be suppressed for strong enough disorder 
see, for instance, Ref.~\cite{Burrell2007}. Assert that the observed 
patterns are due to this requires more research, which is being 
carried out, and depends on the study of the scaling properties of the matrix 
elements of the evolution operator associated with the problem.

Fig. 3  shows that, apparently, 
the system is more robust for smaller transfer times 
than for larger ones, at least for some intervals of the variable $J_{max}$. This is so, in part, because the figure shows the transfer 
probability against a single parameter $J_{max}$ while the detailed behavior of 
$P$ depends on the whole set of exchange couplings. All 
the exchange couplings are bounded by $J_{max}$, but this bound does not give
a detailed information about the distribution of the $J_i$. 
Stochastic optimization methods can be trapped in a "good" minimum and 
do not provide 
information about the landscape topology of the valley where the ``good" 
minimum lies and the width of this valley is what determines the robustness of the transfer 
process against disorder. All this results in that the exchange coupling distributions found for short times are more robust against disorder than the exchange  coupling distributions found for longer times, in certain intervals of the variable $J_{max}$, see Fig. 3 a), b) and c).

 The transferred population $P$ is a probability, but the information that it shows is not enough to quantify large changes in the ECDs. An interesting quantity to study is the probability of transferring an excitation from a given site  of the chain, $m$, to any other site of the chain, $n$, given by $P_{mn}(J_{max})$, where $J_{max}$ is the value  which was used to determine the ECD. It is clear that the TP is given by $P_{1,N}$. Studying the statistical divergence \cite{statistical-divergence} between two probability distributions $D(\lbrace P_{nm}(J_{max})\rbrace, \lbrace P_{nm}(J_{max}'\rbrace)$,  it is possible to determine their properties and how susceptible they are to disorder. We use $\lbrace P_{nm}(J_{max})\rbrace$ to denote the set of probabilities with $1\leq m,n\leq N$ . We have carried out  preliminary studies using the Renyi divergence \cite{renyi-divergence}, $D_{1/2}$ , and  verified that the divergence reproduces the  patterns structure observed in Fig. 3 and Fig. 6.

Fig. 3 shows vertical regions, for instance in Fig. 3 c) there are three clearly appreciable regions whose limits are $J_{max}\leq 5$, $ 5 \leq J_{max}\leq 12$ and $12 \leq J_{max}$. These limits correspond, exactly, with   plateaus observed in the coupling coefficient distributions. In the different regions it can be observed that the susceptibility to disorder is different, as the color map indicates. For example, again in Fig. 3c), it can be seen that the distributions of coupling coefficients  such that $ 5 \leq J_{max}\leq 12$ are more susceptible to disorder than the distributions at both sides. In the $ 5 \leq J_{max}\leq 12$ region the propagation of excitations is strongly suppressed by the increase in the strength of the disorder. In regions where the system is not very susceptible to disorder, the excitation that is initially located at one end of the chain spreads over a large portion of the chain and refocuses at the time of transmission. In areas that are strongly affected by the disorder, the initial  excitation spreads  over a few sites in the chain and the refocusing hardly occurs, which results in  a very poor transferred population. This analysis was carried out calculating the inverse participation ratio, a quantity that is commonly used to analyze transport properties in disordered systems and  quantum state transmission, see Ref.~\cite{Zwick1} and references therein.

The coupling distributions resulting from the application of the PM to chains 
with the Heisenberg Hamiltonian show that it is possible to reach a regime where 
$P = 1-\epsilon$ for chains of any length, not only for chains whose length is 
a power of 2, for arrival times of the order of N as long as the magnitude of 
the couplings is not bounded. At present we cannot affirm whether the regime for 
chains with lengths that are powers of two is the same as for chains of 
arbitrary length. We will  implement more numerical calculations to try to 
elucidate this issue.

{\em Acknowledgments}
The authors acknowledge partial financial  support from
CONICET (PIP11220150100327, PUE22920170100089CO). O.O and P.S. acknowledges 
partial 
financial support from CONICET and SECYT-UNC.

\newpage

\begin{center}
{\large \bf Pretty good quantum state transfer on isotropic and anisotropic
Heisenberg spin chains with tailored site dependent exchange couplings

 Supplemental Material}
\end{center}

\section*{Comparison between the averaged and "worst"
transferred populations}

\begin{figure}[hbt]
\includegraphics[width=0.8\linewidth]{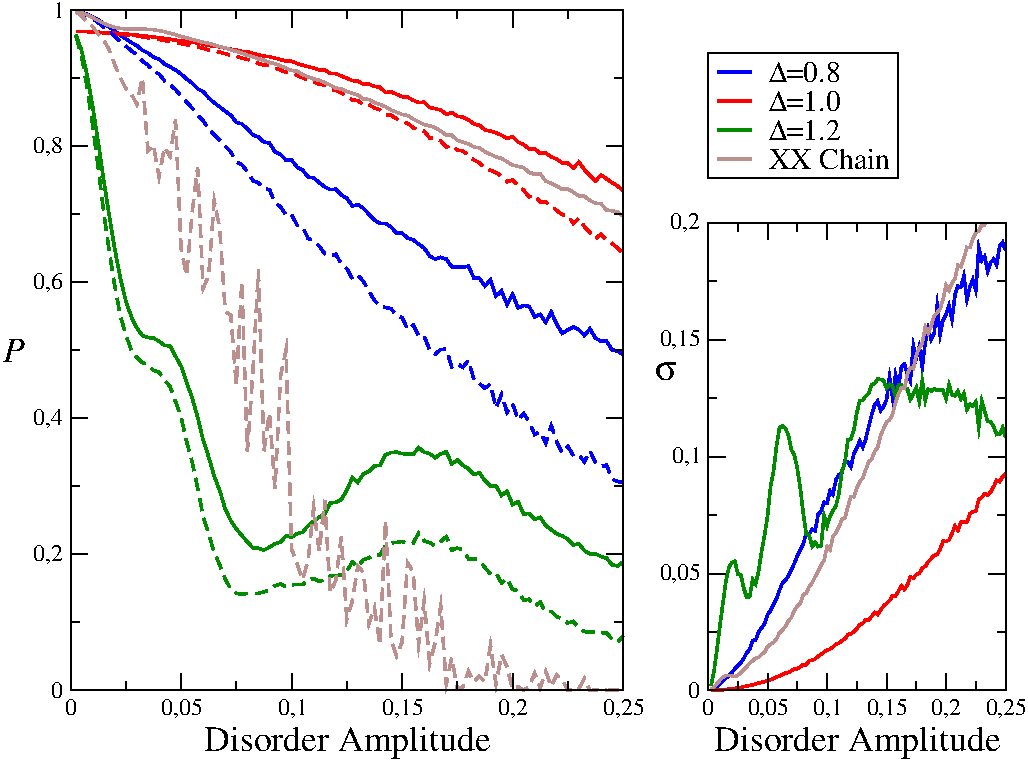}
\caption{ The average transferred population (continuous line)  and the 
"worst case" in a single experiment (dashed line)  versus the strength
of the static disorder. The data shown was obtained for chains of length
$N=51$, $J_{max}=14$ and $N_r=10000$ realizations of the static disorder. See the 
text for details.
}\label{figui}
\end{figure}

The figure above shows the average transferred population versus the strength 
of the static disorder (solid lines) for chains of length $N=51$. The arrival 
time is $T=N$. The red line corresponds to a Heisenberg chain, $\Delta = 1$, 
while the blue and green lines correspond to $\Delta = 0.8$ and $1.2$, respectively, while the brown lines correspond to a spin chain whose dynamics is governed by the XX Hamiltonian, {\em i.e.} $\Delta=0$.
. The dashed lines correspond to the "worst case", that is, the 
minimum population transfer obtainable in a single experiment. It can be 
observed that without static disorder the $\Delta<1$ chain has a better 
performance than $\Delta=1$ and $\Delta>1$, but that for disorder strengths larger 
than 0.05 the Heisenberg chain is clearly superior as a transmission channel, 
being the distance between the average and the "worst case" always smaller 
for that chain.

For comparison reasons we have also included results for an XX chain which are depicted using brown lines, despite that the main interest of the manuscript are chains with XXZ Hamiltonians. It is interesting to note that, with respect to the averaged transferred population, the XX TP shows three different regimes. In the first one its TP is indistinguishable  from the TP obtained for the $\Delta<1$, for disorder amplitudes $a\lesssim 0.05$. For larger values of the disorder amplitude and up to $a\approx 0.075$ the TP of the XX chain remains larger or comparable to the Heisenberg case, while for $a>0.075$ the  TP of the Heisenberg chain is always larger than all the other cases. 

From the results shown for the averaged TP it could be concluded that, if   it is known a priori which is the expected amplitude of disorder and it is possible to implement any type of chain, then the most suitable transmission channels are the XX chain or the Heisenberg chain. Nevertheless, as shown by the worst case curves, the  TP of the XX chain shows by far the worst ``worst case'', see the brown dashed curve in the figure. Moreover, even though that all the curves were obtained using the same number of disorder realizations, it is clear from the large fluctuations shown by the worst case for the XX chain that the statistics of the TP of the XXZ cases is quite different from that of the XX chain. The large fluctuations of the worst case are not as surprising since it is not a good statistical estimator as the standard deviation or the average are. The convergence of the worst case TP to its true value is quite slower in the XX case when compared with the other cases.

The right panel shows the statistical standard deviation of the transferred 
population. The color code is the same as that used in the panel on the left. 
Again it can be seen that the dispersion of the values in the case 
of the Heisenberg chain is less than for the other two chains. If $P_i$ is the 
transferred population obtained for the $i-$th realization of the disorder, 
the averaged transferred population was introduced in the main text
\begin{equation}
 P = \left\langle P\right\rangle_{\xi} = \frac{1}{M} \sum_{i=1}^{N_r} P_i ,
\end{equation}
where $N_r$ is the number of realizations considered, the minimum transferred 
population is given by
\begin{equation}
 P_{min} = \min_i (P_i),
\end{equation}
and the statistical standard deviation of the transferred 
population can be calculated as
\begin{equation}
 \sigma = \frac{1}{N} \sum_{i=1}^{N_r} (P_i -P)^2.
\end{equation}

Here, it is worth to point out that the standard deviation of the XX chain is quite close  to the anisotropic $\Delta<1$ case.

\section*{Exchange coupling distribution for a Heisenberg Chain
 with $N=50$}

In the following we give  a table with the numerical values of all 
the quantities used 
in Figures 5 and  6 of the main text. They correspond to an 
Isotropic Heisenberg model with $N=50$ sites , and transfer times $T=\kappa\,N$,
with $\kappa=1/2,\;1,\;2,\;3,\;4,$ and $5$. 

%\textcolor{blue}{ Caption Figure 1: Value of the largest exchange coupling $J_M$
%searched  within an hypercube of size $J_{max}$ using the pivot method, for a system of $N=30$ spins and  (a) $T=2 N$ ; (b) $T=3N$.}

\onecolumngrid

\begin{table}%[h!]

\begin{center}
{\scriptsize

\begin{tabular}{c|c|c|c|c|c|c}
 $\kappa$     & $1/2$   & 1       &   2     &    3     &  4       &   5 \\  \hline \hline \\
$P$      & 0.9557  & 0.9531  & 0.9550  & 0.9564   &  0.9530  &  0.9573   \\  \hline \\
$J_{max}$&  19.0   &  9.5    &  4.5    & 3.5      &  2.5     &   2.0   \\  \hline \\
$J_1$    & 0.39944 & 0.20749 & 0.10238 & 0.067054 & 0.051485 & 0.041025 \\ \hline \\
$J_2$    &  10.848 & 4.2954  & 1.9585  & 3.3773   & 1.0095   & 0.92646   \\  \hline \\
$J_3$    &  9.6847 & 2.9696  & 2.0284  & 0.60075  & 0.93922  & 0.3956  \\ \hline \\
$J_4$    &  18.273 & 8.5247  & 4.2234  & 2.7508   & 1.413    & 1.2784  \\  \hline \\
$J_5$    &  7.2718 & 7.3141  & 2.4777  & 2.1946   & 1.2422   & 1.7371  \\ \hline \\
$J_6$    & 7.4144  & 4.1991  & 2.2276  & 3.4409   & 1.9408   & 1.7506   \\  \hline \\
$J_7$    & 12.375  & 6.3237  & 2.3133  & 2.2059   & 1.1415   & 1.7628   \\ \hline \\
$J_8$    & 16.283  & 5.2431  & 3.4801  & 2.1465   & 1.1807   & 1.6982   \\ \hline \\
$J_9$    & 14.393  & 6.6373  & 3.3664  & 0.92832  & 2.0173   & 0.42639   \\ \hline \\
$J_{10}$ & 8.6577  & 7.2860  & 3.8836  & 1.4471   & 1.5871   & 1.7395  \\ \hline \\ 
$J_{11}$ & 15.283  & 5.8929  & 2.9566  & 2.2067   & 1.6632   & 1.0526   \\  \hline \\
$J_{12}$ & 15.390  & 7.6673  & 2.7579  & 3.1806   & 1.7769   & 1.9995   \\ \hline \\
$J_{13}$ & 18.846  & 7.1904  & 4.1133  & 3.0605   & 1.9180   & 1.9632  \\  \hline \\
$J_{14}$ & 12.222  & 7.6271  & 4.1504  & 3.3954   & 1.5966   & 1.8553   \\ \hline \\
$J_{15}$ & 17.139  & 7.5777  & 3.8859  & 3.4753   & 1.7934   & 1.9868   \\  \hline \\
$J_{16}$ & 18.045  & 6.8020  & 4.5000  & 3.5000   & 2.1608   & 1.9033   \\ \hline \\
$J_{17}$ & 14.860  & 9.1612  & 3.2385  & 3.5000   & 1.9981   & 1.8423  \\ \hline \\
$J_{18}$ & 18.980  & 6.7268  & 4.5000  & 2.2440   & 2.1971   & 1.7891  \\ \hline \\
$J_{19}$ & 19.000  & 7.9982  & 4.2445  & 3.1035   & 1.6376   & 1.6230  \\ \hline \\
$J_{20}$ & 13.197  & 8.9228  & 3.7993  & 1.9047   & 2.5000   & 1.1124 \\ \hline \\
$J_{21}$ & 19.000  & 7.9298  & 3.9049  & 3.2019   & 1.8421   & 1.6021  \\  \hline \\
$J_{22}$ & 10.844  & 6.5456  & 2.9507  & 1.8041   & 1.8788   & 1.3331    \\ \hline \\
$J_{23}$ & 12.583  & 3.6055  & 3.9359  & 2.8118   & 1.0171   & 0.9925  \\  \hline \\
$J_{24}$ & 13.037  & 5.6579  & 1.4901  & 2.6015   & 0.87489  & 1.3820  \\ \hline \\
$J_{25}$ & 5.645   & 3.7480  & 3.9673  & 1.5246   &2.3294    & 1.6348  \\ \hline
\end{tabular}
}
\caption{\label{tfig5} Values of $P$ and $\{J_i\}$ obtained with the pivot 
method for global optimization inside an hypercube of size $J_{max}$ for $N=50$,  used to made Figures 5 and  6 of the paper.}

\end{center}

\end{table}


\begin{thebibliography}{99}

\bibitem{Nikolopoulos2015}G.M. Nikolopoulos and I. Jex (Edts.), {\em Quantum 
State Transfer and Network Engineering}, Springer-Verlag Berlin Heidelberg 2014

\bibitem{Bose2003}S. Bose, Phys. Rev. Lett. {\bf 91}, 207901 (2003).

\bibitem{Bose-review}S. Bose, Contemporary Physics, 48:1, 13-30 (2007)


\bibitem{Christandl2004}M. Christandl, N. Datta, A. 
Ekert, and A. J. Landahl, Phys. Rev. Lett. {\bf 92},187902 (2004).

\bibitem{Christandl2005}M. Christandl, N. Datta, T. C. 
Dorlas, A. Ekert, A. Kay, and A. J. Landahl, Phys. 
Rev. A {\bf 71}, 032312 (2005).


\bibitem{Zwick1}A. Zwick and O. Osenda, J. Phys. A: Math. Theor. {\bf 44}
105302 (2011).

\bibitem{Zwick2}A. Zwick, G. A. \'Alvarez, J. Stolze, 
and O. Osenda, Phys. Rev. A {\bf 84}, 022311 (2011).

\bibitem{Zwick3}A. Zwick, G. A. \'Alvarez, J. Stolze and  O. 
Osenda, Quantum Information and Computation {\bf 15},  0582 (2015).


\bibitem{Yung2006}Man-Hong Yung, Phys. Rev. A {\bf 74}, 030303(R) (2006).


\bibitem{Burgarth2005}D. Burgarth and S. Bose, Phys. Rev. A {\bf 71}, 
052315 (2005).

\bibitem{Burgarth2005b}D. Burgarth and S. Bose, New Journal of Physics 
{\bf 7},  135 (2005).


\bibitem{Banchi2010}L. Banchi, T. J. G. Apollaro, A. Cuccoli, R. Vaia, 
and P. Verrucchi, Phys. Rev. A {\b 82}, 052321 (2010).

\bibitem{Banchi2011}L. Banchi, T. J. G. Apollaro, A. Cuccoli, R. Vaia
and P. Verrucchi, New J. Phys. {\bf 13}, 123006 (2011).


\bibitem{Wang2010}X. Wang, A. Bayat, S. G. Schirmer, and S. 
Bose, Phys. Rev. A {\bf 81}, 032312 (2010).

\bibitem{Jurdjevic1972}V. Jurdjevic and H. J. Sussmann, J. Diff. Eqn. {\bf 12}, 
313 
(1972).


\bibitem{Burgarth2009}D. Burgarth, S. Bose, C. Bruder, and V. Giovannetti, 
Phys. 
Rev.
A {\bf 79}, 060305(R) (2009).



\bibitem{Ramakrishna1995}V. Ramakrishna, M. V. Salapaka, M. Dahleh, H. Rabitz, 
and
A. Peirce, Phys. Rev. A {\bf 51}, 960 (1995).


\bibitem{Wang2016}X. Wang, D. Burgarth, and S. Schirmer, Phys. 
Rev. A {\bf 94}, 052319 (2016).

\bibitem{Burgarth2010}D. Burgarth, K. Maruyama, M. Murphy, S. 
Montangero, T. Calarco, F. Nori, and M. Plenio, Phys. Rev. A {\bf 81}, 040303(R) 
(2010).

\bibitem{Yang2010} S. Yang, A. Bayat, S. Bose, Phys. Rev. A {\bf82}, 022336 
(2010) .


\bibitem{Heule2010}R. Heule, C. Bruder, D. Burgarth, and V. M. 
Stojanovic, Phys. Rev. A {\bf 82}, 052333 (2010).


\bibitem{Stefanatos2019}D. Stefanatos and E. Paspalakis, Phys. 
Rev. A {\bf 99}, 022327 (2019).



\bibitem{Watanabe2010}G. Watanabe, Phys. Rev. A A {\bf 81}, 021604(R) 
(2010).


\bibitem{Kostak2007}V. Kostak, G. M. Nikolopoulos, and I. Jex, Phys. 
Rev. A {\bf 75}, 042319 (2007)

\bibitem{quantum-dot-chain}D. M. Zajac, T. M. Hazard, X. Mi, E. Nielsen, and 
J. R. Petta, Phys. Rev. Applied {\bf 6}, 054013 (2016).

\bibitem{Li2018}X. Li, Y. Ma, J. Han, Tao Chen, Y. Xu, W. Cai, H. 
Wang, Y.P. Song, Zheng-Yuan Xue,
Zhang-qi Yin, and Luyan Sun, Phys. Rev. Applied {\bf 10}, 054009 (2018)


\bibitem{nuclear-spin-chain}Jingfu Zhang, Gui Lu Long, Wei Zhang,
Zhiwei Deng, Wenzhang Liu, and Zhiheng Lu, Phys. Rev. A {\bf 72}, 012331 
(2005); P. Cappellaro, C. Ramanathan, and D. G. Cory, Phys. Rev. A {\bf 76}, 
032317 (2007); J. Zhang, M. Ditty, D. Burgarth, C. A. Ryan, 
C. M. Chandrashekar, M. Laforest, O. Moussa,
J. Baugh, and R. Laflamme, Phys. Rev. A {\bf 80}, 012316 (2009)

\bibitem{Loft2011}N. J. S. Loft et al  New J. Phys. {\bf 18} 045011 
(2016).

\bibitem{Banchi2011prl}L. Banchi, A. Bayat, P. Verrucchi,
and S. Bose, Phys. Rev. Lett. {\bf 106}, 140501 (2011).

\bibitem{Chapman2016}R. J. Chapman, M. Santandrea, Zixin 
Huang, G. Corrielli, A. Crespi,
Man-Hong Yung, R. Osellame and A. Peruzzo, Nat. Comm.
 {\bf 7}, 11339 (2016).
 
 \bibitem{Kandel2019}Yadav P. Kandel, Haifeng Qiao, Saeed Fallahi, GeoffreyC. 
Gardner, MichaelJ. Manfra, JohnM. Nichol, Nature 573 (2019) 553.

\bibitem{Coden2020}D.S.Acosta Codena, S.S.G\'omez, A.Ferr\'on, O.Osenda, 
Physics Letters A {\bf 387} 127009 (2021).

\bibitem{Zhang2016}X.P. Zhang, B. Shao, S. Hu, J. Zou, L.A. Wu, Ann. Phys. 375 
(2016) 435–443.


\bibitem{Gong2007}J. Gong and P. Brumer, Phys. Rev. A {\bf 75}, 032331 
(2007).


\bibitem{Murphy2010}M. Murphy, S. Montangero, V. Giovannetti, 
T. Calarco, Phys. Rev. A {\bf 82}, 022318 (2010).


\bibitem{Faroq2015}U. Farooq, A. Bayat, S. Mancini, S. Bose, Phys. Rev. B 
{\bf 91},  134303 (2015).
 



\bibitem{Kay2019}Alastair Kay - arxiv-1906.06223 perfect and pretty good state 
transfer for field-free Heisenberg chains 

\bibitem{Kay2010}Alastair Kay, International Journal of Quantum Information
Vol. 8, No. 4 (2010) 641-676

\bibitem{Banchi2017}Leonardo Banchi, Gabriel Coutinho, Chris Godsil, and Simone 
Severini, Journal of Mathematical Physics 58, 032202 (2017)
, 
\bibitem{vanBommel2010}Christopher M. van Bommel, Pretty Good State Transfer 
and Minimal Polynomials, arXiv:2010.06779v1


\bibitem{Zhang2018} Xiao-Ming Zhang, Zi-Wei Cui, Xin
Wang, and Man-Hong Yung, Phys. Rev. A {\bf 97}, 052333 (2018).


%
\bibitem{ssk97} P. Serra, A. F. Stanton and S. Kais,  
Phys. Rev. E {\bf 55}, 1162 (1997).
%
\bibitem{sskb97} P. Serra, A. F. Stanton,  S. Kais and  R. E. Bleil, 
J. Chem. Phys. {\bf 106}, 7170  (1997).
%
\bibitem{numericailrecipes} W.H. Press, S.A. Teukolsky, W.T. Vetterling and B.P. Flannery,
{\it Numerical Recipes in Fortran 77}, 2nd ed., Cambridge University Press,
Cambridge (1996).



\bibitem{Petrosyan2010}D. Petrosyan, G. M. Nikolopoulos, and P. 
Lambropoulos, Phys. Rev. A {\bf 81}, 042307 (2010).




\bibitem{DiFranco2008}C. Di Franco, M- Paternostro and M. S. Kim, Phys. Rev. 
Lett. {\bf 101}, 230502 (2008).

\bibitem{Burrell2007}C- K- Burrell and T. J. Osborne, Phys. Rev. Lett. {\bf 
99}, 167201 (2007). 





\bibitem{Kandel2021}Y. P. Kandel, H. Qiao, and
J. M. Nichol, Appl. Phys. Lett. 119, 030501 (2021). 


\bibitem{Martins2017} F. Martins, F. K. Malinowski, P. D. Nissen, S. Fallahi, 
G. C. Gardner, M. J. Manfra, C. M. Marcus, and F. Kuemmeth, Phys. Rev.
Lett. {\bf 119} 227701 (2017).



\bibitem{statistical-divergence} S. Eguchi, Hiroshima Math. J. 
	{\bf 15} 341 (1985).


\bibitem{renyi-divergence} T. V. Erven, P. Harremoës, 
	IEEE Transactions on Information Theory. {\bf 60}, 3797 (2014).

\bibitem{Nikolopoulos2013} G. M. Nikolopoulos,  Phys. Rev. A 87, 042311 (2013)


\bibitem{sup-mat}Supplemental Material


\end{thebibliography}
\end{document}